\documentclass{einformatyka}
\usepackage{graphicx} %this package allows using pictures, images, etc.
\usepackage[utf8]{inputenc}
\usepackage[sort&compress,numbers]{natbib}
\usepackage{enumitem}
\usepackage[colorinlistoftodos]{todonotes}
\usepackage{color,soul}
\usepackage{multirow}
\usepackage{threeparttable}
\usepackage{xtab}
\usepackage[toc,page]{appendix}
\usepackage{comment}
\usepackage{booktabs}

\begin{document}

% \title{Issues with Search Strings in Systematic Literature Reviews}
\title{How good are my search strings? Reflections on using an existing review as a quasi-gold standard}
%\subtitle{Subtitle of article}

\begin{articleinfo} %this section keeps simple info about the article

\begin{authorgroup} %this section keeps info about article's authors
	\begin{author} %a single author
		\firstname{Huynh Khanh Vi} %firstname
		\surname{Tran}	%surname
		\orgname{Blekinge Institute of Technology} %name of organization (e.g. university) represented by author
		\orgdiv{Department of Software Engineering, SE-37179, Karlskrona, Sweden} %division of organization represented by author
		\email{huynh.khanh.vi.tran@bth.se} %author's e-mail
	\end{author}
	
	%here you can place another authors between \begin{author} and \end{author}
	
	\begin{author}
		\firstname{Jürgen}
		\surname{Börstler}
		\orgname{Blekinge Institute of Technology}
		\orgdiv{Department of Software Engineering, SE-37179, Karlskrona, Sweden}
		\email{jurgen.borstler@bth.se}
	\end{author}

	\begin{author}
		\firstname{Nauman bin}
		\surname{Ali}
		\orgname{Blekinge Institute of Technology}
		\orgdiv{Department of Software Engineering, SE-37179, Karlskrona, Sweden}
		\email{nauman.ali@bth.se}
	\end{author}
	
	\begin{author}
		\firstname{Michael}
		\surname{Unterkalmsteiner}
		\orgname{Blekinge Institute of Technology}
		\orgdiv{Department of Software Engineering, SE-37179, Karlskrona, Sweden}
		\email{michael.unterkalmsteiner@bth.se}
	\end{author}
	
\end{authorgroup}

\keywordset{search string construction, automated search validation, quasi-gold standard, systematic literature review, systematic mapping study} % keyword set allows improved searching

\begin{abstract}[en] %this section contains summary of the article with the information about language [en] or [pl]
\textbf{Background}: Systematic literature studies (SLS) have become a core research methodology in Evidence-based Software Engineering (EBSE).
Search completeness, i.e., finding all relevant papers on the topic of interest, has been recognized as one of the most commonly discussed validity issues of SLSs.\\
\textbf{Aim}: This study aims at raising awareness on the issues related to search string construction and on search validation using a quasi-gold standard (QGS). Furthermore, we aim a%t providing guidelines for search string validation.\\
\textbf{Method}: We use a recently completed tertiary study as a case and complement our findings with the observations from other researchers studying and advancing EBSE.\\
\textbf{Results}: We found that the issue of assessing QGS quality has not seen much attention in the literature, and the validation of automated searches in SLSs could be improved. Hence, we propose to extend the current search validation approach by the additional analysis step of the automated search validation results and provide recommendations for the QGS construction.\\
\textbf{Conclusion}: In this paper, we report on new issues which could affect search completeness in SLSs. Furthermore, the proposed guideline and recommendations could help researchers implement a more reliable search strategy in their SLSs.
\end{abstract}

% % here you can put an epigraph (which is optional)
% \begin{epigraph}
% % To be or not? To be!
% % --Wally Shakelance
% \end{epigraph}

\end{articleinfo}

\section{Introduction}
Systematic literature studies (SLS), including systematic literature reviews, systematic mapping studies and tertiary studies, have become core methods for identifying, assessing, and aggregating research on a topic of interest~\cite{kitchenham_evidence-based_2015}. The need for completeness of search is evident from the quality assessment tools for SLS with questions like: ``was the search adequate?'', ``did the review authors use a comprehensive literature search strategy?'' or ``is the literature search likely to have covered all relevant studies?''~\cite{ali_critical_2019,ali_reliability_2018,Usman2021}. Several guidelines and recommendations have been proposed to improve the coverage of search strategies employed in SLS, e.g., using multiple databases~\cite{kitchenham_evidence-based_2015}, or using an evaluation checklist for assessing the reliability of an automated search strategy~\cite{ali_reliability_2018}. While these guidelines and assessment checklists can be used to design a search string with a higher likelihood of good coverage, these are mostly subjective assessments. 

During the design phase of an SLS, the main instrument researchers have for assessing the likely coverage of their search strings is using a known set of relevant papers that a keyword-based search ought to find~\cite{zhang_identifying_2011, dieste2007developing}. Such a set of known relevant papers is referred to as the quasi-gold standard (QGS) for an SLS. Thus, a QGS is a subset of a hypothetical gold standard, the complete set of all relevant papers on the topic.

Ali and Usman~\cite{ali_reliability_2018} suggest the following for identifying a known set of relevant papers: a) the researchers' background knowledge and awareness of the topic, b) general reading about the topic, c) papers included in related secondary studies, d) using a manual search of selected publication venues.
Kitchenham et al.~\cite{kitchenham_evidence-based_2015} suggest guidelines regarding the size of a QGS for a typical systematic review or a mapping study. The quality of QGS, as a representative sample of the actual population, is critical for deciding how good is a search string.
Nevertheless, the QGS size alone is not sufficient for assessing the QGS quality. 
The diversity of studies in a QGS is also an important quality criterion as it increases the likelihood of being a representative subset of actual related papers. However, to the best of our knowledge, we have not found any related work on validating QGS quality or specific issues relating to using an existing SLS as a source for a QGS.

In a recent tertiary study~\cite{tran2021assessing} on test artifact quality, as suggested by Kitchenham et al.~\cite{kitchenham_evidence-based_2015}, we constructed a QGS by collecting relevant papers from an earlier tertiary study with a related broader topic~\cite{garousi_systematic_2016} (software testing). Our assumption was that a tertiary review of software testing research, in general, would also cover secondary studies on the relatively narrower topic of test artifact quality.

While validating the search in this tertiary study, we have identified issues with the subject area filter in Scopus, the usage of the generic search term ``software'' as a limiting keyword in search, and issues with the search validation approach using a QGS. Based on our experience from constructing and validating search strings using a QGS, we have derived recommendations on validating automated search and constructing the QGS.
Together with the existing guidelines in the literature for the search process, our recommendations help researchers construct a more reliable search strategy in an SLS.

The remainder of the paper is structured as follows: Section~\ref{sec:guidelines_search_validation} provides an overview of guidelines for search validation.
Section~\ref{sec:related_work} presents the related work and our contribution. 
Section~\ref{sec:two_tertiary_studies} summarizes the search process and search validation in our tertiary study~\cite{tran2021assessing}.
Section~\ref{sec:findings} presents our findings when comparing the search results between the two tertiary studies~\cite{tran2021assessing, garousi_systematic_2016}.
Section~\ref{sec:discussion} details the found issues related to search string construction and search validation using QGS.
Section~\ref{sec:proposed_guidelines} presents our proposed guidelines for validating the automated search and constructing the QGS for researchers undertaking large scale SLSs.
Lastly, Section~\ref{sec:concl} concludes the paper.

\section{Guidelines for search validation}\label{sec:guidelines_search_validation}
Several guidelines exist for implementing SLSs with instructions on how to perform the search process~\cite{kitchenham_evidence-based_2015, petersen2015guidelines, ali_critical_2019}.
Kitchenham et al.~\cite{kitchenham_evidence-based_2015} provided detailed instructions on each step of a systematic review procedure.
In particular, regarding the study search process, Kitchenham et al.~\cite{kitchenham_evidence-based_2015} discussed the search completeness concept and different strategies to validate search results.
Accordingly, a search strategy should aim to achieve an acceptable level of search completeness while considering the time constraint and limit in human resources.
Ultimately, the level of completeness depends on the type of the review (qualitative or quantitative)~\cite{kitchenham_evidence-based_2015}.
The completeness could be assessed subjectively based on expert opinion or objectively based on precision and recall~\cite{zhang_identifying_2011, dieste2007developing}.
% There are also measures calculated based on recall and precision values such as F-measure, Matthews correlation coefficient, or Cohen's kappa.
% Nevertheless, these measurements are used more often in machine learning to assess binary classification performance than validating search performance in SLSs.
% In the context of this study, we focus on using recall and precision only.
The recall of a search string, also called sensitivity, is the proportion of all the relevant papers found by the search.
The precision is the proportion of the papers found by the search which are relevant to the study.
By calculating the precision of a search, researchers could estimate the effort required to analyze the search result.

To compute recall and precision, ideally, researchers need to know the number of all relevant papers on the review topic, which is also called the gold standard.
However, it is not easy to acquire the gold standard~\cite{kitchenham_evidence-based_2015,zhang_identifying_2011}, especially when the review domain is not limited.
Hence, a quasi-gold standard, a subset of the gold standard, could be used instead.
There are several approaches listed by Kitchenham et al.~\cite{kitchenham_evidence-based_2015} to acquire a quasi-gold standard.
They include asking experts in the review topic, using a literature review in the same or overlapping topic, conducting an informal automated search, or performing a manual search in specific publication venues within a certain period.
Proposed by Zhang et al.~\cite{zhang_identifying_2011}, the last approach is claimed to be more objective and systematic in assessing automated search results than building the quasi-gold standard based solely on researchers' knowledge.
In general, Zhang et al.'s proposed search strategy could be summarized as follows:
\begin{enumerate}%[noitemsep]
\item Identify publication venues (conferences, journals), databases and search engines. 
The venues are for manual search to establish a quasi-gold standard.
The databases and search engines are for the automated search for relevant papers to answer the research question(s). It is worth noting that the selection of venues is still based on the researchers' domain knowledge; hence, this approach could potentially introduce as much bias as the approach of building a QGS by asking domain experts.
\item Establish the QGS.
The QGS is built by conducting a manual search on the selected publication venues.
All papers published in the given venues within a predefined time frame should be assessed based on the defined inclusion/exclusion criteria.
\item Construct search strings for the automated search.
There are two ways to construct the search strings: (1) based on researchers' domain knowledge and experience; (2) based on word/phrase frequency analysis of the papers in the QGS.
\item Conduct automated search.
The automated search is conducted using the search strings on the selected search engines/databases identified in the previous steps.
\item Evaluate search performance.
The search performance is evaluated based on two criteria, qua\-si-sen\-si\-ti\-vi\-ty (recall) and precision.
Depending on the predefined threshold (70\%--80\% as suggested by Zhang et al.), the search result could be either accepted and merged with the QGS or search strings should be revised until the automated search performance reaches the threshold.
\end{enumerate}

\section{Related work}\label{sec:related_work}

Besides the general guidelines for the search process and search validation described in Section~\ref{sec:guidelines_search_validation}, various issues related to search strategies that could affect the search completeness have been discussed in the literature~\cite{ampatzoglou_identifying_2019, ampatzoglou2020guidelines, bailey_search_2007, dieste2007developing, imtiaz2013tertiary}.
We organized the reported issues into three groups.

The most common issue is the inadequacy of a search strategy in finding relevant publications~\cite{ampatzoglou_identifying_2019, ampatzoglou2020guidelines, bailey_search_2007, dieste2007developing, imtiaz2013tertiary}, which directly affects the search completeness.
Ampatzoglou et al.~\cite{ampatzoglou_identifying_2019, ampatzoglou2020guidelines} discussed the issue via one of their proposed validity categories, namely \textit{study selection validity}.
In this category, the threat ``adequacy of relevant publication''~\cite{ampatzoglou_identifying_2019,ampatzoglou2020guidelines}, which the authors quote, is about ``has your search process adequately identified all relevant primary studies?''.
The authors did not provide further explanation or description of this threat.
Still, they presented a list of mitigation actions such as conducting snowball sampling, conducting pilot searches, selecting known venues, comparing to gold standards.
Based on these mitigation actions, we could see that this validity threat is about whether a search process has identified a representative set of relevant studies.
It is noteworthy that our tertiary study~\cite{tran2021assessing} has applied all their proposed mitigation actions related to this threat except having an external expert review our search process.
Bailey et al.~\cite{bailey_search_2007} conducted three searches on three different topics to analyze the overlaps between search engines in the domain of software engineering.
They reported that the selection of search engines and search terms could influence the number of found papers.
One relevant finding is that for the topic \textit{Software Design Patterns}, their general search terms (``software patterns empirical'' and ``software design patterns study'') offered the highest recall, especially in Google Scholar.
It is worth noting that they define the recall as a percentage of included papers found by a search engine out of the total number of included papers.
To cope with the adequacy of relevant publication in the domain of software engineering experiment, Dieste et al.~\cite{dieste2007developing} discussed the trade-off between high recall and high precision in search.
They proposed criteria for selecting databases and also reported lessons learned when building search terms.
They also noted that using any synonyms of \textit{experiment} alone would omit a huge set of relevant papers when searching articles reporting software engineering experiments.
Imitiaz et al.~\cite{imtiaz2013tertiary}, in their tertiary study, discussed different issues which could affect the adequacy of relevant publication in SLRs.
These issues are search terms with many synonyms and unknown alternatives, the trade-off between generic and specific search string, search approaches (automated, manual, snowball sampling) selection, search level (title, abstract, keywords) and abstract quality.

The second most common issues which could impact the search completeness are inconsistencies and limitations of search engines and databases~\cite{kruger_search_2019, bailey_search_2007, chen2010towards,ali_reliability_2018}.
Bailey et al.~\cite{bailey_search_2007} identified two main issues with search engines: inconsistent user interfaces and limitations of search result display.
They concluded that search engines do not provide good support for conducting SLRs due to these two issues.
The inconsistencies in databases and search engines' internal search procedures and their output are also reported by Ali and Usman~\cite{ali_reliability_2018} and by Kr\"uger et al.~\cite{kruger_search_2019}.
As reported in Kr\"uger et al.'s study~\cite{kruger_search_2019}, API search results in databases could vary even within the same day.
On top of that, databases and search engines evolve over time, which could lead to changes in their search API~\cite{ali_reliability_2018,kruger_search_2019}.
Due to the identified limitations, the selection of search engines and databases becomes essential as it could impact search completeness.
Chen et al.~\cite{chen2010towards} proposed three metrics (overall contribution, overlap, exclusive contribution) to characterize search engines and databases which they called electronic data sources (EDS).
These metrics could help researchers to choose EDS for their literature reviews.
According to the authors, the \textit{overall contribution}, which is about the percentage of relevant papers returned by an EDS, is the dominant factor in selecting EDS.
Meanwhile, the \textit{exclusive contribution} is about papers that could be found by one EDS only.
This information helps researchers to decide which EDS could be omitted.
The \textit{overlap metric} (the papers returned by multiples EDS) could be used to determine the order of EDS in the search process. 
% Another potential way to mitigate this issue is that researchers could save the search results for working offline.
% The saved search strings and search results could then be shared with other researchers for searches comparison and analysis in the future.

The third most common issue is search terms standardization in software engineering~\cite{bailey_search_2007, zhou2016map}.
Bailey et al.~\cite{bailey_search_2007} pointed out that there is a lack of standardization of terms used in software engineering, which could influence the search result adequacy.
They raised the need to have up-to-date keywords and synonyms to mitigate the risk of missing relevant papers.
This standardization issue has also been reported by Zhou et al.~\cite{zhou2016map} as one of the main validity threats in SLRs.

In summary, we have found several studies that reported issues with the search process and the importance of adequate search string construction and validation to achieve search completeness.
In a tertiary study~\cite{tran2021assessing}, we have encountered all of these issues and applied different strategies to mitigate validity threats related to the search process.
These include systematically constructing search strings, piloting searches, selecting well-known digital search engines and databases, and using a relevant tertiary study's search results to build a QGS for search validation.
Nevertheless, we have not identified any related work discussing the quality assessment of QGSs or issues related to the construction of QGSs from existing SLSs.
Hence, based on our experience with evaluating the searches using the QGS, we propose guidelines for automated search and QGS validation, which could help researchers construct a more reliable search strategy in SLSs.

\section{Analysis of using another SLS as QGS}
\label{sec:two_tertiary_studies}
This study is based on two recent tertiary studies~\cite{garousi_systematic_2016, tran2021assessing} conducted independently.
Both articles were published in the Journal of Information and Software Technology.
The first study~\cite{garousi_systematic_2016} on software testing was undertaken by Garousi and Mäntylä, while the second one~\cite{tran2021assessing} with a narrower topic, test artifact quality, was published five years later by the authors of this study.
A high-level overview of both tertiary studies can be found in Table~\ref{tab:study_overviews}.
For convenience, we refer to the tertiary study~\cite{garousi_systematic_2016} on software testing as the ST study and the tertiary study on test artifact quality~\cite{tran2021assessing} as the TAQ study in this paper.

In the TAQ study~\cite{tran2021assessing}, to evaluate the search performance, we constructed a QGS by extracting relevant papers from the ST study~\cite{garousi_systematic_2016}.
A summary of the resulting search strategy and search evaluation outcomes is illustrated in Figure~\ref{fig:search_steps}.
More details about the search process and the search evaluation using QGS are presented in Section~\ref{sec:our_search_process} and Section~\ref{sec:search_evaluation} respectively.
To better understand the result of the search performance evaluation, we also analyzed the differences in search results between the two tertiary studies.
The analysis of these differences is described in Section~\ref{sec:findings}.

\begin{table}
%\arraystretch{1.4}
{    
    \footnotesize
    \begin{center}
    \caption{Overview of Tran et al.'s \cite{tran2021assessing} and Garousi and Mäntylä's \cite{garousi_systematic_2016} tertiary studies.}
    \label{tab:study_overviews}
    \begin{tabular}{p{0.2\textwidth} p{0.36\textwidth}p{0.36\textwidth}} 
    \toprule
    & \textbf{Tran et al. \cite{tran2021assessing}} \textit{(TAQ study)} & \textbf{Garousi and Mäntylä \cite{garousi_systematic_2016}} \textit{(ST study)} \\
    \midrule
    Title & Assessing test artifact quality---A tertiary study & systematic literature review of literature reviews in software testing \\ [4pt]
    Publication year & 2021 & 2016 \\ [4pt]
    Focus & Quality attributes of test artifacts and their measurement. & Mapping of research in software testing. \\ [4pt]
    Research goals &
    To investigate how test artifact quality has been reported in secondary studies in the following aspects: (1) quality attributes of test artifacts; (2) quality measurements of test artifacts; (3) testing-specific context where the quality attributes and quality measurements have been studied. &
    To provide a systematic mapping of secondary studies in software engineering to investigate the following aspects: (1) different areas in software testing; (2) research questions types; (3) demographics of secondary studies; (4) citation of secondary studies. \\ [4pt]
    Automated search & Yes & Yes \\ [4pt]
    Snowballing & No & Yes \\ [4pt]
    Search String & Iterative, see Figure~\ref{fig:search_strings_diff} for details & See Figure~\ref{fig:search_strings_diff} for details \\ [4pt]
    Include SLRs \& SMS & Yes & Yes \\ [4pt]
    Include other reviews/ surveys & No & Yes \\
    \bottomrule
    \end{tabular}
    \end{center}
}
\vspace{5mm}
\end{table}

\subsection{Search process}\label{sec:our_search_process}

\begin{figure*}%[htb]
\begin{center}

\includegraphics[width=1\textwidth]{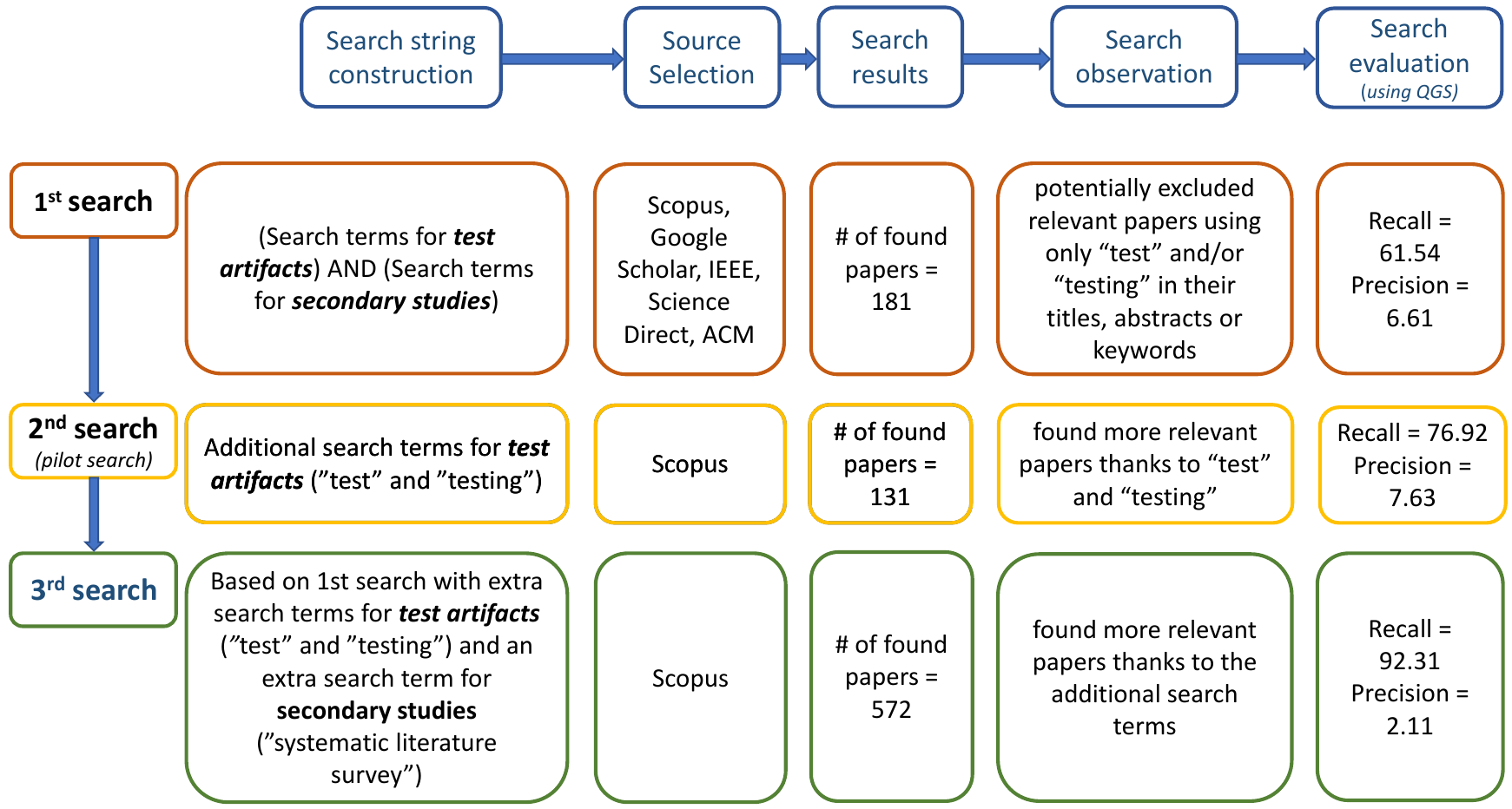}

\caption{Overview of the search steps in the tertiary study on test artifact quality (TAQ study)~\cite{tran2021assessing}.}
\label{fig:search_steps}
\end{center}
\end{figure*}

\begin{figure*}[htb]
\begin{center}
\includegraphics[width=1\textwidth]{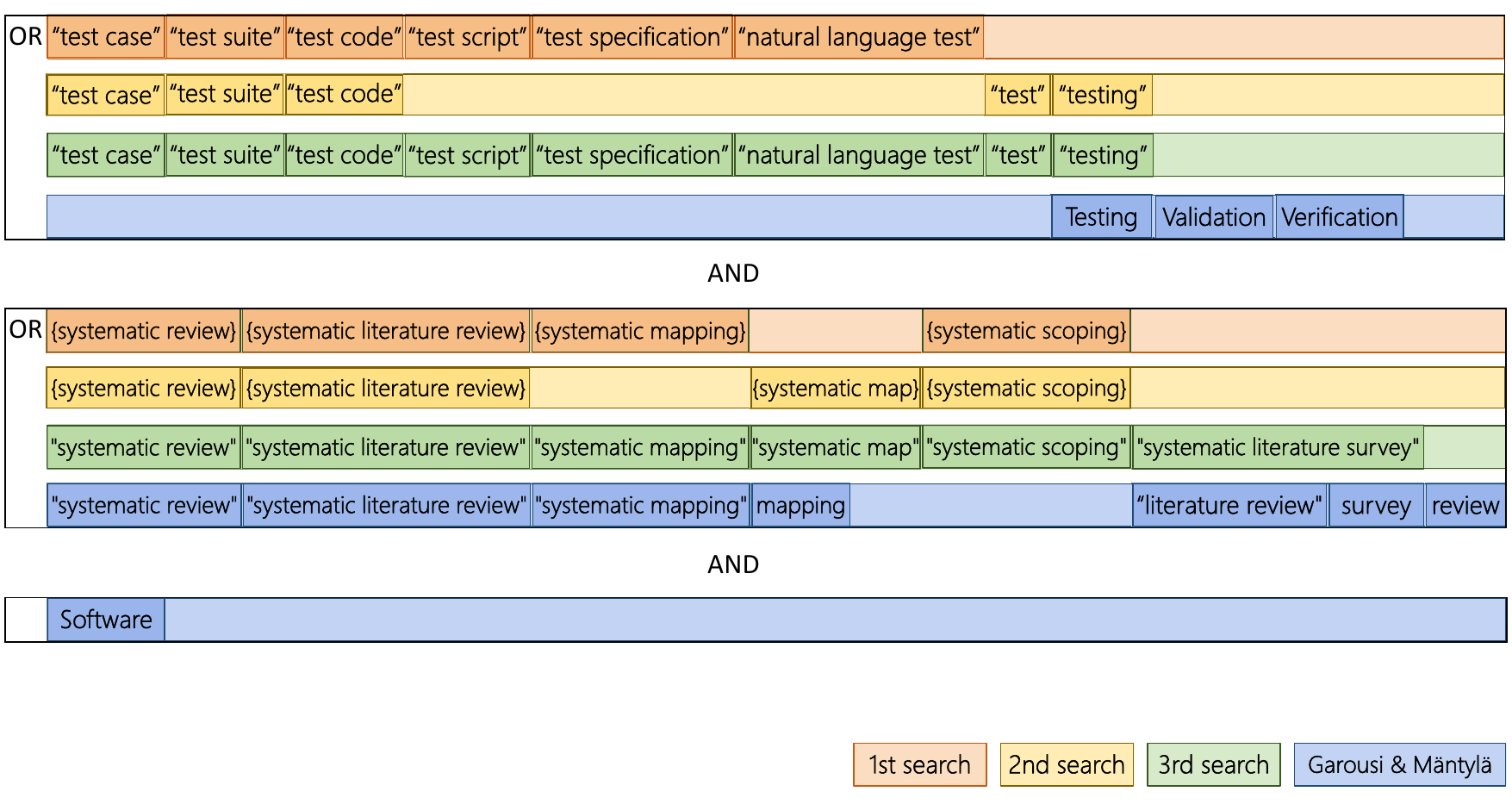}
\caption{Comparison of the search terms used in the search strings of the two tertiary studies, the TAQ study~\cite{tran2021assessing} and the ST study~\cite{garousi_systematic_2016}.}
\label{fig:search_strings_diff}
\end{center}
\end{figure*}

In the TAQ study, test artifact refers to test case, test suite, test script, test code, test specification, and natural language test.
The overview of the study's three searches is illustrated in Figure~\ref{fig:search_steps}, and the search results are presented in Table~\ref{tab:search_results}.
We used a visual analysis tool~\cite{heberle2015interactivenn} called InteractiVenn~\footnote{\url{http://www.interactivenn.net/}} to analyze the overlaps in the search results.
The TAQ study's search terms and their differences with the ST study's search terms are shown in Figure~\ref{fig:search_strings_diff}.

Since the TAQ study's search goal was to identify systematic secondary studies discussing test artifact quality, the search strings needed to capture two aspects: (A) systematic secondary studies and (B) test artifact quality.
Hence, the search strings were constructed as (A AND B).
To address aspect B (test artifact quality), we included search terms to describe test artifact such as ``test case'', ``test script'' while excluding the search term ``quality'' as this latter search term is too common to be useful as a separate component of a search string.

The first search was conducted in April 2019 and returned 181 papers (see Table~\ref{tab:search_results}).
The initial set of 58 SLRs/SMSs found by the ST study was used to validate the completeness of the searches (explanation on how these 58 papers were collected is in Section~\ref{sec:search_evaluation}).
Hence, to verify if the first search was adequate, we screened the titles and abstracts of the 39 SLRs/SMSs, which were not found by the first search but by the ST study only.

Among the 39 SLRs/SMSs, several are on different topics such as software product line testing, testing of web services, mutation testing techniques, etc.
These papers used ``test'' and ``testing'' but no term for test artifact in titles and abstracts.
Since these papers could potentially discuss test artifact quality but were not found by the first search, we considered it as a potential issue of the first search.
In other words, the first search might exclude relevant papers having ``test'' or ``testing'' but no term for test artifact in their titles, abstracts or keywords.

To verify the above hypothesis, we conducted a second search, which is a pilot search in Scopus in October 2019, including the additional search terms ``test'' and ``testing''.
As a result, the second search returned 131 papers (see Table~\ref{tab:search_results}), which contained more relevant papers than the first search.
Hence, we added the additional search terms ``test'' and ``testing'' in the third search to reduce the risk of missing relevant papers. 
Also, the third search included another search term, ``systematic literature survey'', which was inspired by the ST study's search terms.
In other words, the third search was built based on the first search and the confirmed hypothesis from the second search (pilot search).
The third search was conducted in Scopus in October 2019 and restricted to one subject area, ``Computer Science'', to reduce the search noise. 
The third search returned 572 papers, as shown in Table~\ref{tab:search_results}.

\begin{figure}
\begin{center}

\includegraphics[width=0.6\textwidth]{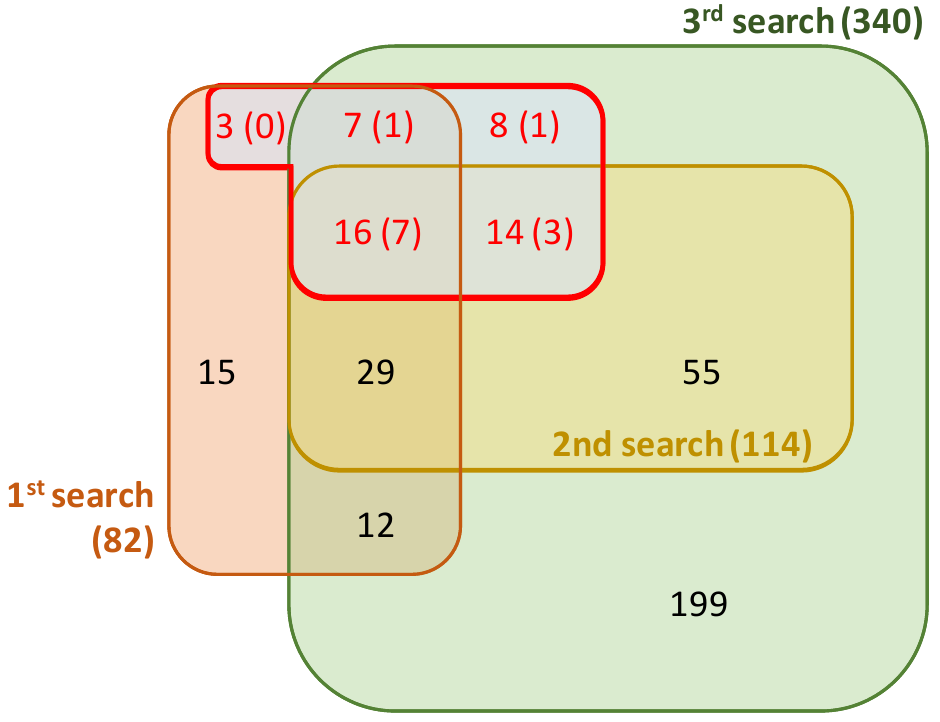}

\caption{Overlaps between three searches in the tertiary study on test artifact quality (TAQ study)~\cite{tran2021assessing}. The red box illustrates the distribution of the selected papers among searches, and the numbers in parentheses show the number of papers belonging to the QGS.}
\label{fig:three_searches_overlap}
\end{center}
\end{figure}

The overlaps between the search results are presented in Figure~\ref{fig:three_searches_overlap}.
All the numbers in the figure refer to papers after deletion of duplicates and obviously irrelevant papers, i.e., papers that are not about software engineering or computer science based on their titles, abstracts and keywords.
The red box shows the distribution of 48 out of the complete set of 49 selected papers among the searches.
One of the 49 selected papers was extracted from the ST study' search result (the decision on selecting papers from the ST study' search result is explained in Section~\ref{sec:search_evaluation}); hence, it is not shown in the figure.

As shown in Figure~\ref{fig:three_searches_overlap}, out of the 82 papers returned by the first search, 8 (1+7) papers were included in the QGS, and 26 (3+7+16) eventually turned out relevant.
By considering the first search and the third search only (since the second search result is a subset of the third search result), the third search returned 276 (8+14+55+199) additional papers, of which a further 4 (1+3) were included in the QGS, and a further 22 (8+14) turned out as relevant.
% Hence, although more effort is needed to screen the additional papers given by the third search, we almost doubled the number of relevant papers with the third search.
Based on the above observation, we could see that most of the QGS papers were found by the first and third search (in total, 12 out of 13 QGS papers).
It also turned out that we almost doubled the number of relevant papers with the third search.
Therefore, we consider including the first and third search as a fair trade-off for this study in terms of the effort required to read papers and the returned benefit (identified relevant papers plus QGS papers). 
Nevertheless, the trade-off between recall and precision could be different depending on the goal of the targeting SLS.
For example, if researchers aim to compare different techniques in software engineering, a high recall might be more desired than a high precision~\cite{kitchenham_evidence-based_2015}.

% As we can see in Figure~\ref{fig:three_searches_overlap}, out of 82 papers returned by the first search, 26 (3+7+16) were relevant.
% The second search gave us 69 (14+55) additional papers, of which 14 were relevant.
% The third search gave us another 207 (8+199) additional papers, of which only 8 were relevant.
% If we consider the first search and the third search only (since the second search result is a subset of the third search result), then the third search returned 276 (8+14+55+199) additional papers, of which 22 papers (14+8) were relevant.
% In that scenario, although more effort is needed to screen the additional papers given by the third search, we almost doubled the number of relevant papers with the third search.
% On top of that, the first search and third search identified twelve out of 13 papers belonging to the QGS as shown in Figure~\ref{fig:three_searches_overlap}.
% Hence, we consider including the first and third search is a fair trade-off for this study in terms of the effort required to read papers and the returned benefit in terms of identified relevant papers and papers from the QGS. 
% Nevertheless, the trade-off between recall and precision could be different depending on the goal of the targeting SLS.
% For example, if researchers aim to compare different techniques in software engineering, a high recall might be more desired than a high precision~\cite{kitchenham_evidence-based_2015}.
% To ensure that the searches are sufficient for data extraction, we evaluated the searches' performance using a QGS described in Section~\ref{sec:search_evaluation}.

\begin{comment}
\begin{table}
{    \footnotesize
    \begin{center}
    \caption{Search terms for search strings}
    \label{tab:search_terms}
    \begin{tabular}{p{0.1\textwidth}p{0.35\textwidth}p{0.4\textwidth}}
        \toprule
        \textbf{Search} & \textbf{Substring A -- Test artifact} & \textbf{Substring B -- Secondary studies}\\
        \midrule
        1st search & ``test case'' OR ``test suite'' OR ``test script'' OR ``test code'' OR ``test specification'' OR ``natural language test'' & ``systematic review'' OR ``systematic literature review'' OR ``systematic mapping'' OR ``systematic scoping'' \\
        2nd search & ``test case'' OR ``test suite'' OR ``test code'' OR ``test'' OR ``testing'' & ``systematic review'' OR ``systematic literature review'' OR ``systematic map'' OR ``systematic scoping'' \\
        3rd search & ``test case'' OR ``test suite'' OR ``test script'' OR ``test code'' OR ``test specification'' OR ``natural language test'' OR ``test'' OR ``testing''  & ``systematic review'' OR ``systematic literature review'' OR ``systematic mapping'' OR ``systematic map'' OR ``systematic scoping'' OR ``systematic literature survey'' \\
        \bottomrule
    \end{tabular}
    \end{center}
}
\end{table}
\end{comment}

\begin{table}
{    
    \footnotesize
    \begin{center}
    \caption{Search Results of the tertiary study on test artifact quality (TAQ study)~\cite{tran2021assessing}}
    \label{tab:search_results}
    \begin{tabular}{p{0.08\textwidth}p{0.38\textwidth}p{0.12\textwidth}p{0.25\textwidth}} 
        \toprule
        \textbf{Search} & \textbf{Database/ \mbox{Search Engine}} & \textbf{\# of \mbox{papers}} & \textbf{Search Level} \\
        \midrule
        \multirow{8}{*}{1st} & Scopus & 100 & Title, abstract, keywords \\
        & Google Scholar & 27 & Title \\
        & IEEE & 16 & Title, abstract, keywords \\
        & Science Direct & 23 & Title, abstract, keywords \\
        & ACM & 15 & Title, abstract \\
        \cmidrule{2-3}
        & Total & 181 & \\
        & Excl. duplicates & 121 & \\
        & Excl. duplicates \& clearly irrelevant studies & 82 & \\ [4pt]
        \midrule
        \multirow{3}{*}{2nd} & Scopus & 131 & Title \\
        & Excl. duplicates & 131 & \\
        & Excl. duplicates \& clearly irrelevant studies & 114 \\
        \midrule
        \multirow{3}{*}{3rd} & Scopus & 572 & Title, abstract, keywords \\
        & Excl. duplicates & 569 & \\
        & Excl. duplicates \& clearly irrelevant studies & 340 \\
        \bottomrule
    \end{tabular}
    \end{center}
}
\vspace{5mm}
\end{table}

\subsection{Search performance evaluation using a QGS}\label{sec:search_evaluation}
In this section, we describe how a QGS was constructed in the TAQ study.
We then explain how the recall and precision of the first and third searches in this tertiary study were computed based on the QGS.
In this evaluation process, we focused on the first search and third search only as the second search was actually a pilot search, and its result is a subset of the third search's (more details in Section~\ref{sec:our_search_process}).

It is worth emphasizing that we did not follow the instructions for constructing the QGS given by Zhang et al.~\cite{zhang_identifying_2011} (more details on their instructions could be found in Section~\ref{sec:guidelines_search_validation}).
Overall, the key difference is that we extracted relevant papers from the ST study~\cite{garousi_systematic_2016} to build the QGS, while Zhang et al. suggested constructing a QGS by conducting a manual search in some publication venues with a specific time span.
Our decision on how to construct the QGS is motivated by the fact that the ST study is a peer-reviewed tertiary study conducted by the domain experts and its topic (software testing) is related to and broader than the TAQ study's topic (test artifact quality).
Using another literature review to collect known relevant papers for search validation is also one of the suggestions by Kitchenham et al.~\cite{kitchenham_evidence-based_2015}.

It is also necessary to mention that, although there is no information regarding the complete set of found papers, the ST study has provided access to its initial set of 123 papers which is the result of the ST study's authors removing clearly irrelevant papers from their search result~\cite{garousi_systematic_2016}.
By analyzing the 123 papers, we found two duplicate papers (having the same title, authors and abstract).
Of the remaining 121 papers, 63 are \textit{informal/regular} surveys, i.e., reviews without research questions as stated in the ST study.
Hence, we focused on the remaining 58 (121 - 63) papers, which are SLRs/SMSs as the TAQ study considered systematic reviews and mappings only. 

When considering all the 58 SLRs/SMSs papers from the initial set of papers in the ST study~\cite{garousi_systematic_2016} as the QGS, the first and third searches found 18 and 44 papers from the QGS, respectively.
The recall and precision of the two searches are relatively low, as shown in Table~\ref{tab:recall_precision}.
Since these 58 papers might contain irrelevant papers to the scope of the TAQ study, we updated the QGS with the 13 relevant papers from the set of 58 SLRs/SMSs papers.
The 13 papers were selected according to the TAQ study's study selection criteria (explained in Appendix~\ref{appendix:studySelectionCriteria}).

The distribution of the updated QGS over the first and third searches is shown in Figure~\ref{fig:twoSearches_GMSearch_QGS}.
We need to note that all numbers in Figure~\ref{fig:twoSearches_GMSearch_QGS} refer to papers after deletion of duplicates, obviously irrelevant papers and informal reviews.
On the one hand, the two searches' precision decreased as the number of QGS papers found by the searches decreased (from 18 and 44 to 8 and 12 papers by the first and third search respectively).
On the other hand, with this more accurate QGS, the recall of the two searches increased by a significant margin.
Also, as shown in Table~\ref{tab:recall_precision}, even though the third search returns a higher reading load than the first search, it is still superior to the first one in terms of identifying relevant papers.

We considered two directions at this point: (1) select relevant papers from the first and third search for data extraction; or (2) do forward snowball sampling on the 13 relevant papers found by the ST study then select relevant papers from there.
To pick an appropriate direction, we first conducted a first-step forward snowball sampling in Scopus on the 13 papers and calculated its recall and precision using the relevant papers found by the first search only as the QGS.
We found 946 papers citing the 13 papers.
The set reduced to 832 papers after removing duplicates (same title, abstracts, and authors).
This set of 832 papers includes the ST study itself.
Among these 832 papers, 10 of them met the TAQ study's study selection criteria (explained in Appendix~\ref{appendix:studySelectionCriteria}).
Since the 13 papers were published between 2009 and 2015, our assumption was that forward snowball sampling on these 13 papers should help us identify relevant papers published from 2009 onward.
Hence, we selected the 20 relevant papers published from 2009 found by the first search but not by the ST study as the QGS.
As shown in Table~\ref{tab:recall_precision}, the recall and precision of the forward snowball sample were much lower than the ones of the third search.
We might have found more relevant papers and improved the recall if conducting a more extended snowball sampling on the 13 papers.
However, considering the low possibility of getting a higher recall than the third search and yet much more effort required for the more extended snowball sampling, we decided to use the results of the first and third search and the initial set of 58 SLRs/SMSs papers from the ST study for the paper selection.

\begin{table}
{    
    \footnotesize
    \begin{center}
    \caption{Recall and Precision of searches in the tertiary study on test artifact quality (TAQ study)~\cite{tran2021assessing}.}
    \label{tab:recall_precision}
    \begin{tabular}{p{0.15\textwidth}p{0.3\textwidth}p{0.3\textwidth}}
        \toprule
        \multicolumn{3}{l}{\textbf{Considering all 58 SLRs and SMSs from the ST study's initial set as the QGS}} \\
        % \multicolumn{3}{l}{\textbf{found by Garousi and Mäntylä as the QGS}} \\
        \midrule
        & 1st search & 3rd search \\
        Recall & 32.76 & 75.86 \\
        Precision & 15.70 & 7.73 \\
        \bottomrule
        \toprule
        \multicolumn{3}{l}{\textbf{Considering only the 13 relevant SLRs and SMSs from the ST study's initial set as the QGS}} \\
        \multicolumn{3}{l}{\textbf{(see also the last column in Figure~\ref{fig:search_steps})}} \\
        \midrule
        & 1st search & 3rd search \\
        Recall & 61.54 & 92.31 \\
        Precision & 6.61 & 2.11 \\
        \bottomrule \toprule
        \multicolumn{3}{l}{\textbf{Considering the 20 relevant SLRs and SMSs found by the 1st search but not by the ST study}} \\
        % \multicolumn{3}{l}{\textbf{found by our 1st search but not}} \\
        \multicolumn{3}{l}{\textbf{as the QGS}} \\
        \midrule
        & First-step forward snowballing & \\
        Recall & 50.00 & \\
        Precision & 1.20 & \\
        \bottomrule
    \end{tabular}
    \end{center}
}
\vspace{5mm}
\end{table}

\section{Findings}\label{sec:findings}
While evaluating the performance of the first and third searches in the TAQ study (described in Section~\ref{sec:search_evaluation}), we also analyzed the differences in search results between the evaluated searches and the ST study's search.
The purpose of the search result comparison is to understand better why the searches in the TAQ study achieve certain recall and precision and if these searches have any issues that we could fix or mitigate to improve their recall and precision.
In this section, we report our findings from this search results comparison.
The overlaps in search results between the two tertiary studies are shown in Figure~\ref{fig:twoSearches_GMSearch_QGS}.
 
Regarding the ST study's search result, there are two things we need to remark.
First, in this search result comparison, the ST study's search result refers to its initial set of 58 SLRs/SMSs.
These 58 papers do not contain informal/regular surveys, duplicate or clearly irrelevant papers to their study's topic (software testing) (more details on how these 58 papers were collected are in Section~\ref{sec:search_evaluation}).
Hence, before comparing the search results, we also removed duplicate and clearly irrelevant papers found in the first and third searches. 
As a result, there were 82 and 340 remaining papers, respectively, from the first and third search.
Second, there is no information regarding when the ST study concluded its search.
As the latest publication date of the papers found by the ST study's search is October 2015, we assume that the search found papers published until October 2015.

\subsection{The first search and the ST study's search}\label{sec:1stSearch_Garousi}
As shown in Figure~\ref{fig:twoSearches_GMSearch_QGS}, the first search found 63 (16+47) papers not included in the ST study's search result.
Among those 63 papers, 26 papers were published before October 2015, which are within their assumed search period.
The first possible explanation is that the first search included five search engines and databases (see Table~\ref{tab:search_results}), while the ST study searched on Scopus and Google Scholar.
Indeed, six out of those 26 papers are from ACM and Science Direct.
Second, the first search did not include the search term ``software'', which was mandatory in the ST study's search.
Due to this difference in the search string construction, out of the 26 papers, the first search found 11 more papers.
One interesting note is that the remaining nine papers (26-6-11) could be found when applying the ST study's search string on Scopus and Google Scholar.
It is possible that those papers were not indexed by Scopus or Google Scholar by the time the ST study's search was conducted.

The ST study found 39 papers (4+23+1+11) (as  shown in Figure~\ref{fig:twoSearches_GMSearch_QGS}) that were not included in the first search's result.
Among these 39 papers, 33 of them did not have any terms for \textit{test artifact} in title, abstract and keywords which is required by the first search.
The remaining six papers (39-33) did not use the term ``systematic'' in title, abstract and keywords; hence, they were also excluded by the first search, which only looked for systematic reviews.

\subsection{The third search and the ST study's search}\label{sec:3rdSearch_GarousiSearch}
As shown in Figure~\ref{fig:twoSearches_GMSearch_QGS}, the third search found 296 (249+47) papers that were not in the ST study's search result.
Among these 296 papers, the first search found 47 of them. 
The possible reasons for the ST study's search result not containing those 47 papers are explained in Section~\ref{sec:1stSearch_Garousi}.
For the remaining 249 (297-47) papers, 84 were published before October 2015, which meets their assumed search period.
Out of these 84 papers, 31 did not use the term \textit{software} in their titles, abstracts or keywords, which is one of the required search terms of the ST study.
However, the other 53 papers (84-31) meet the ST study's search string.
We suspect that Scopus did not index these 53 papers by the time the ST study conducted its search.

The ST study found 14 papers (2+1+11) (as shown in Figure~\ref{fig:twoSearches_GMSearch_QGS}) which the third search's result did not include.
Six out of the 14 papers were not peer-reviewed; hence, they are out of the scope of this comparison.
Among the other eight papers (14-6) which were peer-reviewed, three of them did not use ``systematic'' in their titles, abstracts or keywords, and two of them~\cite{barbosa2011software,sharma2021testing} were included under the subject area ``Engineering'' in Scopus.
The third search did not find these five papers as the search accepted only systematic reviews and was limited to the subject area ``Computer Science'' in Scopus.
The other three papers (8-3-2) are not indexed in Scopus but other search engines/databases (Google Scholar, INSPEC, ACM), and two of them were found by the first search, which included those databases and search engines.
We discuss the differences between the two searches next.

\subsection{The first search and the third search}\label{sec:1stSearch_3rdSearch}
\begin{figure}%[htb]
\begin{center}

\includegraphics[width=0.7\textwidth]{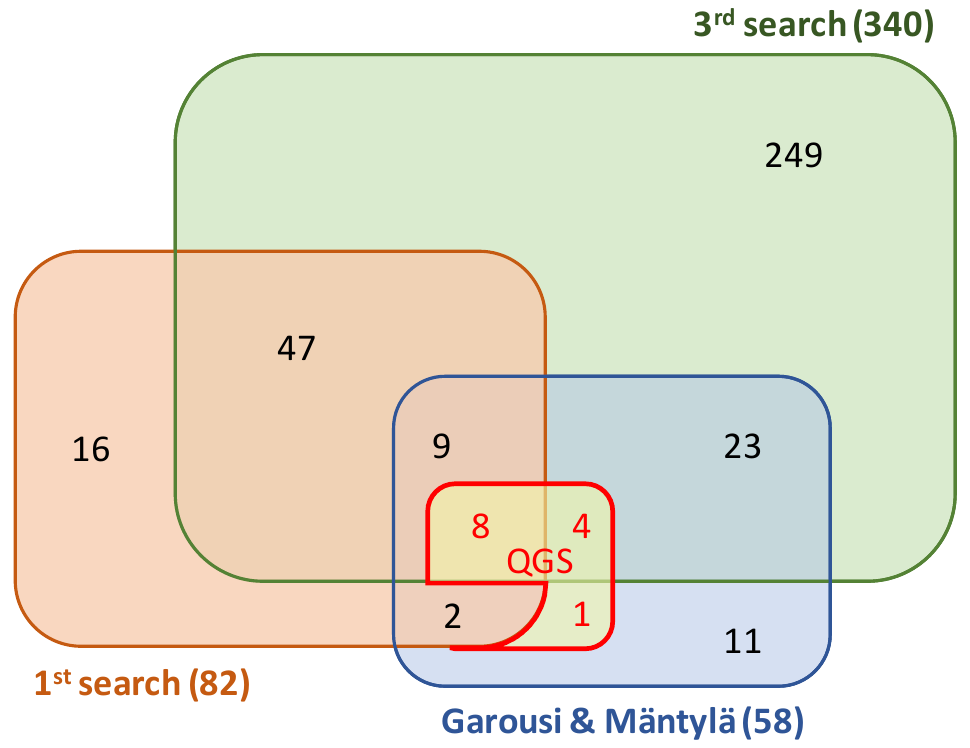}

\caption{Overlaps between the first and third searches and the 58 SLRs/SMSs papers from the initial set of papers in the ST study~\cite{garousi_systematic_2016}. The red box illustrates the distribution of the papers of the QGS.}
\label{fig:twoSearches_GMSearch_QGS}
\end{center}
\end{figure}

The first search found 18 papers (16+2 as  shown in Figure~\ref{fig:twoSearches_GMSearch_QGS}) which the third search's result did not contain.
Among those 18 papers, five of them were not categorized under the subject area ``Computer Science'' but different subject areas (three papers~\cite{paul2012redefinition,munasinghe2016supply,ahmad2017systematic} under ``Engineering''; one paper~\cite{pradhan2019coverage} under ``Business, Management and Accounting/Decision Sciences/Social Sciences''; and one paper~\cite{arora2018systematic} under ``Multidisciplinary'').
The other 13 papers (18-5) were found in other databases/search engines by the first search (six papers in Google Scholar, four papers in ACM, one paper each in IEEE, Wiley, and Web of Science).
Hence, the main reasons are the databases/search engines selection and the subject area(s) selection in Scopus.

The third search found 276 papers (249+23+4) (as shown in Figure~\ref{fig:twoSearches_GMSearch_QGS}) which the first search missed.
It could be due to the more inclusive search strategy of the third search as it had extra search terms (``test'', ``testing'', and ``systematic literature survey'', as shown in Figure~\ref{fig:search_strings_diff}).

\section{Discussion}\label{sec:discussion}
In this section, we first discuss issues relating to search string construction, then issues relating to using a QGS for search evaluation that we have discovered while evaluating the searches' performance in the TAQ study~\cite{tran2021assessing}.

\subsection{Issues in search string construction}
Based on our findings described in Section~\ref{sec:findings}, we identified the following issues with search string construction in SLSs.

The first issue is about using generic search terms in SLSs.
Based on the differences in search results between the TAQ study~\cite{tran2021assessing} and the ST study~\cite{garousi_systematic_2016}, we found that adding generic terms (\textit{software} in the case of the TAQ study) with the Boolean operator \textbf{AND} to a search string increases the risk of missing relevant papers.
The problem is that in research areas where certain contexts are assumed, some keywords might not be explicitly stated since they are implied.
It is the term \textit{software} in the case of research in software development/quality/engineering.
Hence, ``AND software'' just narrows down the search result as not all papers in software engineering use the term \textit{software} in title-abstract-keywords.
This also supports our decision of not including ``AND quality'' to the search strings.
Oppositely, if generic terms are added to search strings with the Boolean operator \textbf{OR}, researchers likely have more noise in their search results.
We, therefore, regard ``AND software'' and ``AND quality'' as unnecessary excluders due to their threat of excluding relevant papers, while we consider ``OR software'' an unnecessary includer due to its risk of retrieving non-relevant material.

The second issue we have identified is about search filters in Scopus.
Search filters can be applied to various meta-data of a publication, such as language, document type, publication year, etc.
By using search filters, researchers can limit their search results, for example, to papers written in English and published in the year 2021 only.
In the case of the TAQ study case, we focus on the subject area filter in Scopus.
We found that some papers were not categorized correctly according to their subject areas.
For example, the ST study found two papers~\cite{barbosa2011software,sharma2021testing} that could not be found by the third search (as discussed in Section~\ref{sec:3rdSearch_GarousiSearch}).
These papers were classified under the subject area \textit{Engineering} instead of \textit{Computer Science}.
Likewise, the first search found five papers~\cite{paul2012redefinition,munasinghe2016supply,ahmad2017systematic,pradhan2019coverage,arora2018systematic} that were not found by the third search.
These five papers were classified wrongly in different subject areas (see Section~\ref{sec:1stSearch_3rdSearch}) instead of \textit{Computer Science}.
This misclassification could be origin in the algorithm for detecting papers' subject areas in Scopus, inappropriate classification and keywording by the papers' authors, or a combination of both.

The third issue is search repeatability.
We could not replicate the search result by the ST study in Scopus using their search string.
The search repeatability issue has been well discussed in the literature~\cite{kitchenham_repeatability_2011,ampatzoglou2020guidelines,kruger_search_2019,zhou2016map,ali_reliability_2018}.
We referred to the checklist proposed by Ali and Usman~\cite{ali_reliability_2018} for evaluating the search reliability of the ST study's search process.
As a result, we found that some details the ST study could have reported to increase their search repeatability.
Those details include search period, database-specific search strings, additional filters, deviation from the general search string, and database-specific search hits.
The missing information and the potential inconsistencies in the API search of the search engine (Scopus in this case) could be the reasons for issues in search repeatability.

\subsection{Issues related to using Quasi-gold standards}\label{sec:issues_with_QGS}
We have identified two issues related to using a quasi-gold standard (QGS) for search validation.

The first issue is about the QGS characteristics.
To the best of our knowledge, several aspects have not been discussed sufficiently in the literature~\cite{kitchenham_evidence-based_2015, zhang_identifying_2011}.
Kitchenham et al.~\cite{kitchenham_evidence-based_2015} described different approaches to constructing a QGS followed by a discussion on QGS size.
Zhang et al.~\cite{zhang_identifying_2011} proposed a detailed guideline on building a QGS using a manual search on specific publication venues for a certain time span.
We argue that QGS size is not the only aspect on which researchers should focus. We discuss this further and propose some suggestions to overcome this issue in Section \ref{sec:QGS_characteristics}.

The second issue with using the QGS for search validation is about the quality of the QGS itself. By its nature, the QGS is only an approximation of a complete set of relevant papers. However, by conducting more than one search, we could triangulate issues in the QGS and make informed decisions about modifying our search string. 
Comparing our search results to the ST study's search result (the basis for our QGS), we could identify the root causes for not finding certain relevant papers included in the QGS.
This helped us establish whether our searches were simply not good enough with respect to the QGS or whether there were acceptable reasons for missing a paper.
Additionally, the search result comparisons helped us to understand why the QGS did not contain certain relevant papers found by our searches. Thus, it allows us to identify shortcomings of the QGS and have more confidence in the quality of the QGS than relying solely on the recall and precision results.

\section{Recommendations for QGS construction and search validation}\label{sec:proposed_guidelines}
As discussed in Section~\ref{sec:issues_with_QGS}, we argue that recall and precision are important for assessing a search result but that they should not be the only criteria. It is also critical to analyze the root causes for not finding papers that the search should have found by looking into those papers of the QGS that the search missed.
It might turn out that these papers did not use any of the search terms in the title, abstract or keywords or that they used different terminologies. The search can then be modified to ensure that one or more of those papers can be found. However, which root causes are addressed (and how) depends on the potential return on investment, i.e., the number of additional relevant papers that may potentially be found in relation to the total increase in the size of the search result.
We recommend playing through various scenarios and assessing their potential return on investment with the help of precision and recall.

To address the root causes originating in the QGS, we first describe the desirable characteristics of a good QGS in Section~\ref{sec:QGS_characteristics} and then propose recommendations for constructing a QGS in Section~\ref{sec:QGS_construction}.
For root causes originating in the obtained search results, we propose an additional analysis step in Section~\ref{sec:guideline_auto_search}. 
These suggestions are based on our findings (reported in Section~\ref{sec:findings} and Section~\ref{sec:discussion}) when evaluating the searches' performance in the TAQ study~\cite{tran2021assessing}.

\subsection{QGS desirable characteristics}\label{sec:QGS_characteristics}

Fundamentally, a QGS needs to be a good ``representative'' of the gold standard, and having a good QGS is vital for search validation in SLSs.
In this section, we describe desirable characteristics of a good QGS.
The characteristics are based on our experience from using QGS~\cite{AliETMMHKV19,tran2021assessing} and Wohlin et al.'s~\cite{wohlin_reliability_2013} discussion on search as a sampling activity when the entire population (i.e. the set of all relevant papers) is unknown. 
Moreover, we draw inspiration from the snowball sampling guidelines for a good initial set to propose recommendations for arriving at a good QGS~\cite{wohlin2014guidelines}. 

The main characteristics of a QGS discussed in the SE literature are \textit{relevance} and \textit{size}~\cite{kitchenham_evidence-based_2015,zhang_identifying_2011}. For example, Kitchenham et al.~\cite{kitchenham_evidence-based_2015} suggest indications for acceptable QGS sizes for various SLS types. However, as it is impossible to have true gold standards for most SLSs in SE~\cite{zhang_identifying_2011} and the overall population of relevant papers is unknown~\cite{wohlin_reliability_2013}, we argue that size alone is insufficient as an indicator of the quality of a QGS. We, therefore, introduce a third desirable characteristic, diversity, and present the complete list of QGS desirable characteristics as follows:

\begin{enumerate}
    \item \textbf{Relevance}.\\
    Each paper in the QGS should be relevant to the targeted topic.
    Any paper added to the QGS should meet the inclusion criteria of the ongoing SLS. 
    In the TAQ study~\cite{tran2021assessing}, we used the selected papers from a related SLS as a QGS after confirming that those papers met the selection criteria of the study.
    
    \item \textbf{Size}.\\ 
    Unlike \textit{relevance} and \textit{diversity}, where general suggestions have been provided, giving a recommendation for the size for a QGS is difficult since the ``target population'' is unknown. 
    The number of relevant papers for an SLS can vary widely.
    The SLSs in Kitchenham et al.'s SLR of SLRs in software engineering~\cite{kitchenham_systematic_2009} included 6--1485 relevant papers with a median of 30.5 papers. 
    The tertiary study by da Silva et al.~\cite{da_silva_six_2011} lists a range from 4--691 (median: 46). 
    Since the focus of an SLS can be general or narrow, depending on the topic of interest and the type of research questions, providing a general recommendation for the minimal size of a QGS seems impossible.

    \item \textbf{Diversity}.\\
    Diversity entails that a good QGS should comprise papers extracted from independent clusters representing different research communities, publishers, publication years and authors. 
    This is important as even a large, and relevant QGS will be ineffective to objectively assess a search strategy if it is limited in its coverage. 

\end{enumerate}

\subsection{QGS construction}\label{sec:QGS_construction}
There are neither fixed thresholds for quality indicators nor a deterministic way of arriving at a good QGS. However, the following recommendations\footnote{The recommendations in Section~\ref{sec:QGS_construction} are a synthesis of existing guidelines \cite{ali_reliability_2018,ampatzoglou2020guidelines,kitchenham_evidence-based_2015,wohlin2014guidelines} and our own experience as reported in this study and from using QGS in other systematic literature reviews~\cite{AliETMMHKV19}.} for identifying and selecting suitable papers for inclusion in a QGS provide heuristics that will increase the likelihood of creating a diverse QGS that can help determining `is my search good enough' more objectively.   
\begin{enumerate}
   \item \textbf{Identification:}
   There are several approaches researchers could consider to locate relevant papers for their QGS construction:
    \begin{itemize}
    \item Conduct manual search. Researchers first manually identify relevant venues (conferences, workshops, and journals) and researchers. 
    After that, researchers can manually search for relevant papers by reading titles of papers in the selected venues (most common sources are Google Scholar, Scopus, DBLP) and of the selected authors. 
   
    \item Conduct informal search in electronic data sources. 
    We recommend that persons conducting the informal search should be independent researchers. An independent researcher here is not involved in the study and has not participated in the design of the search strategy for the study. We recommend these additional considerations because the search terms used in the informal search might compromise the effectiveness of the QGS as a validation mechanism. 
    For example, if the same search terms are used for the informal search and the actual systematic search, then the recall is likely 100\% since the actual search will probably find the same relevant papers as the informal search but not more than that. Hence, the 100\% recall cannot guarantee that researchers achieve an acceptable level of search completeness.
    We further recommend that researchers should use citation databases like Scopus and Google Scholar in this step to avoid publisher bias.
 
    \item Use expert's recommendation.
    Researchers could have an expert in the field (not involved in the search strategy design) recommend papers for a QGS for the current study. The experts should have access to the research questions and the selection criteria of the study.

    \item Use an existing SLS.
    An existing SLS could be selected as a source of papers for the QGS. 
    Since existing SLSs have been peer-reviewed, and their study selections typically have been validated, researchers will save time using this approach compared to the above approaches.
    However, the topics of existing SLSs will usually differ at least slightly from the topic of the new SLS (otherwise, a new SLS would not be necessary). 
    The QGS might, therefore, not cover the research questions in the new SLS sufficiently.
    Hence, researchers should critically review the search and selection strategy of the selected SLS. We recommend using the checklist provided by Ali and Usman~\cite{ali_reliability_2018} to assist this evaluation. If the SLS had major weaknesses in search, we suggest supplementing the construction of the QGS with the above approaches. 
  
    \end{itemize}
 
    \item \textbf{Selection:}
    The researchers should evaluate the potentially relevant papers identified through the above sources for relevance. We suggest using the selection criteria of the targeted SLS to select papers that should comprise the QGS.

    \item \textbf{When to stop:} 
    An exhaustive search of the potential sources listed above is impractical. After all, this is not the actual search but rather an attempt to create a good validation set for the search strategy. We, therefore, recommend that consulting a combination of sources and selection should be done iteratively until a sufficiently large, relevant and diverse QGS is obtained. What is sufficiently large will depend on the research questions and the breadth of the target research area. Due to the reasons discussed above (in Section~\ref{sec:QGS_characteristics}), we do not recommend any range here and leave it to the subjective judgment of the researchers.
    Nevertheless, we argue that the more diverse a research area is, the larger a QGS is needed.
    As an indication of size, researchers should investigate the numbers of selected studies in existing SLSs in the area or the sizes of QGSs in related SLSs.
    Furthermore, if the QGS will be split for both search string formation and validation, a larger QGS will be required.
    Overall, a good QGS should be diverse, not too small, and relevant for answering the research questions. Primarily, the resulting QGS should have papers from different research communities, publishers, publication years, and authors. 
\end{enumerate}

\subsection{Additional recommendation for search validation using QGSs}\label{sec:guideline_auto_search}
Kitchenham et al.~\cite{kitchenham_evidence-based_2015} have discussed two approaches to validate a search strategy via search completeness (more details in Section~\ref{sec:related_work}). Researchers could use the personal judgment of experienced researchers to evaluate the search completeness.
Since this approach is subjective, it might be challenging to quantify the search completeness level.
The other approach is to measure the completeness level by calculating the precision and recall of searches based on a pre-established QGS.
With the second approach, the completeness assessment becomes objective within the limits of the quality of the QGS. 
This means that the quality assessment of the search string is connected with the quality assessment of the QGS.
In other words, if the QGS was not constructed properly, even a high recall cannot guarantee that the search result is good.

Following the above guidelines will increase our confidence in the precision and recall values. 
While meeting certain search recall and precision thresholds (see~\cite{zhang_identifying_2011,kitchenham_evidence-based_2015}) are necessary, it is also essential to understand how the search achieves these recall and precision scores.
Hence, we suggest researchers perform the additional step of analyzing the differences between the search results and the QGS.
This allows researchers to identify reasons for missing relevant papers with the automated search that are included in the connected QGS, and consequently improve their search strategy or document the limitations. 
For example, we found that it is necessary to be aware that subject areas categorization in some search engines might not categorize papers adequately. When comparing the search results with the QGS, we noticed that we could not find several papers as they were assigned to the wrong categories.

To facilitate this additional step, we suggest that researchers should use tools to analyse the search overlap. The metadata in search results is not consistently formatted across various data sources and often has minor differences like inconsistent capitalization and differences in encoding of special characters. Therefore, care must be taken to clean the data. Reference management tools like Zotero\footnote{Zotero, a free and open-source reference management tool \url{https://www.zotero.org/}} or EndNote\footnote{EndNote, a commercial reference management tool \url{https://endnote.com/}} can be used to compare the search results. Furthermore, the use of visualizations like Figures~\ref{fig:three_searches_overlap} and \ref{fig:twoSearches_GMSearch_QGS} helps to get a better understanding of comparative performance of various search strings.
There are tools that can assist researchers in analyzing and visualizing lists intersections, such as one developed by the Bioinformatics and Evolutionary Genomics Group\footnote{\url{http://bioinformatics.psb.ugent.be/webtools/Venn/}} or InteractiVenn\footnote{\url{http://www.interactivenn.net/}} by Heberle et al.~\cite{heberle2015interactivenn} that we used in this study.

\subsection{Potential limitations}\label{sec:limitations}
The recommendations and additional search validation steps proposed in this study are closely based on our experience while performing automated database searches in a tertiary study on test artifact quality~\cite{tran2021assessing}.
In this tertiary study, we used another related tertiary study~\cite{garousi_systematic_2016} to construct a QGS for the search validation.
There could be other issues if we had used another search strategy or a different QGS construction approach. Therefore, the list of issues is not exhaustive, and the recommendations in this paper may need to be supplemented further.
% Hence, to address search completeness in an SLS, besides our recommendations and guidelines, researchers should also refer to other existing guidelines and suggestions which are applicable for their search strategy.
% Another limitations of the recommendations proposed in this paper is that these are not applicable for SLSs where snowball sampling is used as the main search strategy.

For example, our recommendations for search validation using QGSs might not apply for SLSs with the traditional snowball sampling approach, i.e., all known relevant papers are used as the initial set.
In other words, the QGS and the initial set are the same.
% The reason is that the same set of known relevant papers will be used for both searching papers and validating search results. 
Hence, the recall will always be 100\% but not useful to validate the search completeness.
However, the recommendations could become applicable if researchers split the whole set of known relevant papers into two subsets. 
In this case, one subset of known relevant papers will be used as the initial set for snowballing search, while the other will be used to validate the snowballing search results as the QGS.

\section{Conclusions and Lessons Learned}\label{sec:concl}
Search incompleteness, i.e., the absence of relevant papers in the results produced by the employed search strategy, has been recognized as one of the most commonly discussed validity threats of systematic literature studies (SLSs).
This study reports our experience with mitigating this validity threat while performing searches in a tertiary study on test artifact quality~\cite{tran2021assessing}.
We constructed a quasi-gold standard (QGS) by extracting relevant papers from another relevant tertiary study~\cite{garousi_systematic_2016} published several years before ours.
While evaluating the tertiary study's searches using the QGS, we have found new issues with the search string construction and the search validation approach using a QGS.
The issues could affect search completeness in SLSs.
They relate to using generic search terms with the Boolean operator \textbf{AND}, the subject area filter in Scopus, and the QGS quality.

Consequently, we proposed extending the current search validation approach by the analysis step of the automated search validation results as well as recommendations on the QGS construction.
The main argument of the analysis step of the search validation results is that recall and precision is not enough to validate an automated search. 
Researchers should analyze reasons for the automated search to miss relevant papers included in the QGS.
Likewise, addressing the concern of QGS quality that has not been well studied in the literature, our recommendations on the QGS construction step helps researchers construct a high-quality QGS, i.e., a good ``representative'' of the gold standard.
Ultimately, the extended guideline and recommendations can support researchers achieve a more reliable search process.
To validate and improve the extended guidelines for search validation, we will collect feedback from the software engineering research community via interviews and surveys.

\section*{Acknowledgment}
This work has been supported by ELLIIT, a Strategic Area within IT and Mobile Communications, funded by the Swedish Government. The work has also been supported by research grant for the VITS project (reference number 20180127) from the Knowledge Foundation in Sweden.

\bibliography{mybib}
\bibliographystyle{abbrv} 

\newpage

\begin{appendices}
\section{Study Selection Criteria}
\label{appendix:studySelectionCriteria}

Our study selection inclusion/exclusion criteria are described as follows:

\setlist{nolistsep}
\begin{enumerate}[noitemsep]
\item Phase 1: applied on authors, title and abstract
\begin{itemize}
    \item Exclude papers that: 
    \begin{enumerate}[label=(E\arabic*), noitemsep]
        \item are duplicate papers;
        \item are not systematic studies\footnote{Garousi and Mäntylä's~\cite{garousi_systematic_2016} initial set of 121 papers contained 63 informal surveys without research questions. Since we only were targeting systematic studies, these were excluded.};
        \item are not peer reviewed;
        \item are outside computer science or software engineering.
    \end{enumerate}
\end{itemize}

\item Phase 2: applied on title and abstract
\begin{itemize}
    \item Exclude papers that: 
    \begin{enumerate}[label=(E\arabic*), noitemsep]
        \setcounter{enumii}{4}
        %\item are duplicate papers; (already taken care of)
        \item are not about software testing.
    \end{enumerate}
    \item Include papers that fulfil all of the following:
    \begin{enumerate}[label=(I\arabic*), noitemsep]
        \item are systematic literature reviews (SLR), quasi-SLRs, Multi-vocal literature reviews, or systematic mappings;
        \item discussed or potentially discussed quality of test artifacts
        %(test case, test suite, test script, test code, test specification, natural language test)
    \end{enumerate}
\end{itemize}

\item Phase 3: applied on full text
\begin{itemize}
    \item Exclude studies that:
    \begin{enumerate}[label=(E\arabic*), noitemsep]
        \setcounter{enumii}{5}
        \item Are duplicate studies (two different studies using the same data)
    \end{enumerate} 
    \item Include studies which discussed any of the following:
    \begin{enumerate}[label=(I\arabic*), noitemsep]
        \setcounter{enumii}{2}
        \item definition of the quality of test artifacts;
        \item quality characteristics of test artifacts;
        \item quality attributes of test artifacts;
        \item quality metrics of test artifacts;
        \item tools, methods, approaches, frameworks to assess test artifacts' quality;
        \item guidelines, checklists to write test artifacts.
    \end{enumerate}
\end{itemize}
\end{enumerate}
\end{appendices}

\end{document}